\documentclass[twocolumn,english,prb,showpacs,superscriptaddress]{revtex4-1}
\usepackage[colorlinks=true,urlcolor=blue,citecolor=blue,linkcolor=blue]{hyperref}
\usepackage[sort&compress]{natbib}
\usepackage[pdftex]{graphicx}
\usepackage[T1]{fontenc}
\usepackage[latin9]{inputenc}
\usepackage{amssymb}
\usepackage{graphicx}
\usepackage{amsmath,color}
\usepackage{mathrsfs}
\usepackage{float}
\usepackage{indentfirst}
\usepackage{babel}
\usepackage{bbold}
\usepackage{wasysym}

\begin{document}

\title{First order topological phase transition of the Haldane--Hubbard model}

\author{Jakub Imri\v{s}ka}
\affiliation{Theoretische Physik, ETH Zurich, 8093 Zurich, Switzerland}
\author{Lei Wang}
\affiliation{Theoretische Physik, ETH Zurich, 8093 Zurich, Switzerland}
\affiliation{Beijing National Laboratory for Condensed Matter Physics, and Institute of Physics, Chinese Academy of Sciences, Beijing 100190, China}
\author{Matthias Troyer}
\affiliation{Theoretische Physik, ETH Zurich, 8093 Zurich, Switzerland}

\begin{abstract}
We study the interplay of topological band structure and conventional magnetic long-range order in spinful Haldane model with onsite repulsive interaction. Using the dynamical cluster approximation with clusters of up to 24 sites we find evidence of a first order phase transition from a Chern insulator at weak coupling to a topologically trivial antiferromagnetic insulator at strong coupling. These results call into question a previously found intermediate state with coexisting topological character and antiferromagnetic long-range order.
Experimentally measurable signatures of the first order transition include hysteretic behavior of the double occupancy, single-particle excitation gap and nearest neighbor spin-spin correlations. This  first order transition is contrasted with a continuous phase transition from the conventional band insulator to the antiferromagnetic insulator in the ionic Hubbard model on the honeycomb lattice.
\end{abstract}



\maketitle

\section{Introduction}
The Haldane model~\cite{Haldane88} describes non-interacting fermions on a honeycomb lattice in a staggered magnetic field. Over the past decade, this 
prototypical model of a topologically non-trivial bandstructure  has inspired numerous developments in the field of topological insulators,~\cite{Hasan:2010ku, Qi:2011hb} and has recently been experimentally realized using ultracold fermions in an optical lattice.~\cite{HaldaneRealization14} Because of their high degree of controllability, ultracold atomic gases offer a unique opportunity to investigate the interplay of topological bandstructure and the strong interactions, where one expects a variety of fascinating phenomena.\cite{Hohenadler:2013dx}

To experimentally investigate the interaction effects on the Haldane model, one loads two species of ultracold fermionic atoms into an optical lattice and tunes their onsite interaction. 
However, the Haldane--Hubbard model poses a theoretical challenge. The lack of  time-reversal symmetry gives rise to a severe fermion sign problem~\cite{SignProblem} and limits the use of quantum Monte Carlo (QMC) methods.~\cite{SignProblemInQMC}
This is in contrast with the time-reversal symmetric Kane--Mele--Hubbard (KMH) model, in which the two spin species experience opposite magnetic flux. The KMH model thus allows sign-problem free QMC simulations at half filling that show a continuous phase transition from the quantum spin Hall insulator into an antiferromagnetic insulator (AFI) as the interaction strength increases.\cite{KMH11,KMHbyQMC13,PiKMH14} 

Similarly, in the Haldane--Hubbard model the local onsite interaction favors an AFI in the strong coupling regime,~\cite{PhysRevB.91.134414, HaldaneHubbard_ED_DMRG_15} which competes with a Chern insulator (CI) at weak coupling. To find out how the two limiting cases are connected requires a non-perturbative treatment. Being hard to tackle, some of the previous studies used static mean-field approximations.~\cite{SlaveBoson, PhysRevB.84.035127, HaldaneHubbardMeanField_2015, PhysRevB.83.205116, HaldaneHubbardMeanField_2015b} All these studies reported an additional phase with coexisting antiferromagnetic long-range order and non-trivial topological character at intermediate interaction strengths.  This topologically non-trivial AFI state has a clear mean-field picture: in the vicinity of a putative second order quantum phase transition to the AFI, the antiferromagnetic order parameter increases continuously so that there is a finite region where the topological band gap persists despite of the counteracting topologically trivial band gap due to the magnetic order. However, given the approximate nature of the static mean-field treatment, it is hard to assess whether this intermediate state really exists. 

In this paper we thus study the ground state phase diagram of the Haldane--Hubbard model using the dynamical cluster approximation~(DCA),~\cite{DCA2001,QClusterTheoriesReview2005} which is a cluster extension of dynamical mean-field theory (DMFT).\cite{DMFTReview96} By using clusters embedded in a self-consistently determined bath, both short-range correlations within the cluster and long-range correlations are captured. 
Solving embedded clusters with up to 24 sites at low temperature 
we can go beyond static mean-field and exact diagonalization treatments. Our main result is a \emph{first order} phase transition from a topologically non-trivial band insulating state to a magnetic long-range ordered state, preempting the intermediate ``topological AFI'' state. Observables such as the antiferromagnetic magnetization, double occupancy, all exhibit hysteretic behavior around the transition point, which are clear signatures of a first order phase transition.\cite{HysteresisInOpticalLattice_Esslinger_15}

\section{Model and Method}\label{sec:model_method}
\begin{figure}
\centering
\includegraphics[width=8.5cm]{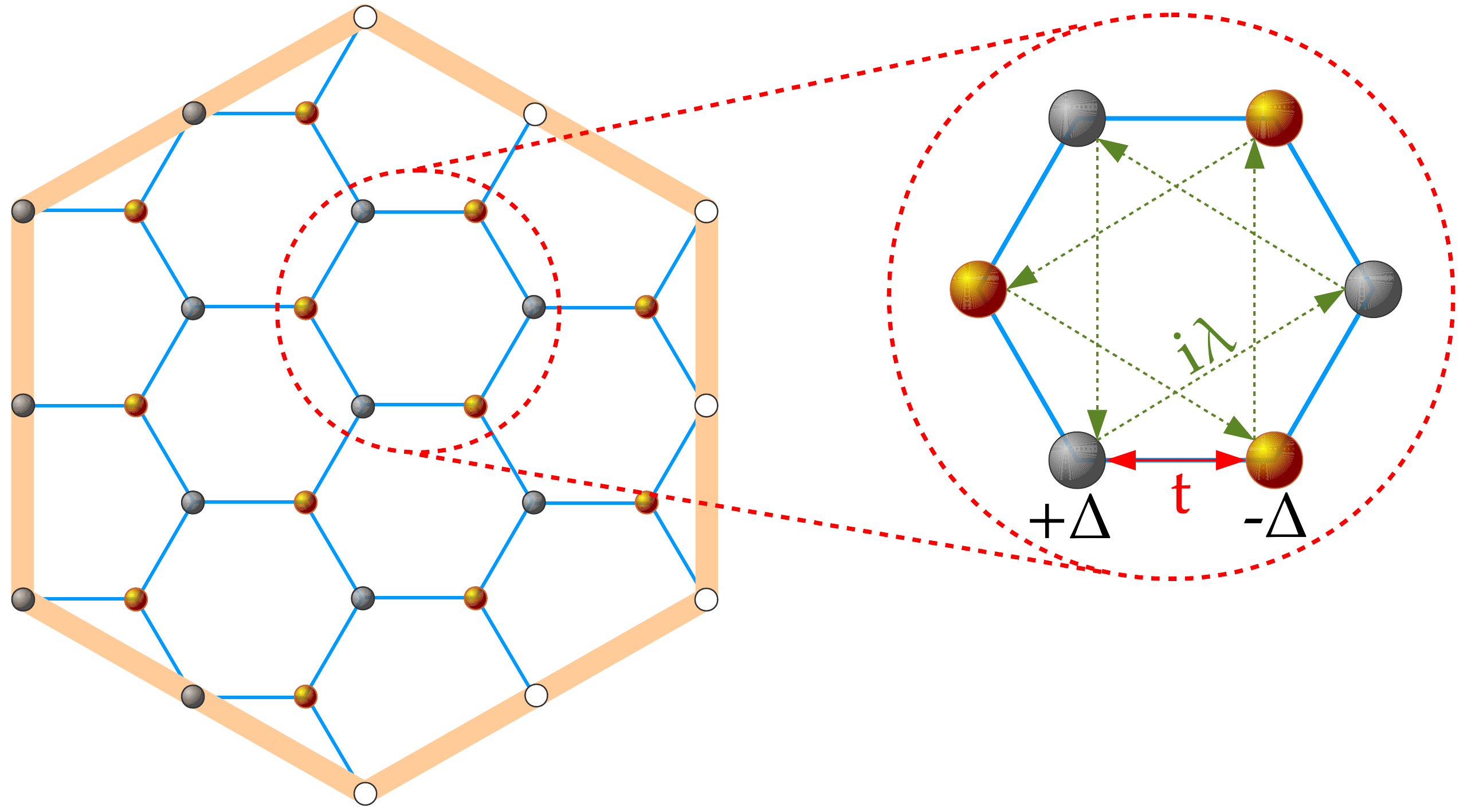}
\caption{The embedded cluster with 24 sites used to obtain the phase diagram. White sites on the border correspond via periodic boundary conditions to gray sites on the opposite border. The enlarged hexagon shows the terms of the non-interacting part of the Hamiltonian Eq.~(\ref{eq:Ham}): nearest neighbor hopping $t$, next nearest neighbor hopping $\mathrm{i}\lambda$, and staggered potential, which is $+\Delta$ ($-\Delta$) on the sublattice with orange (gray) sites.}
\label{fig:model}
\end{figure}

The Hamiltonian of the Haldane--Hubbard model reads
\begin{eqnarray}
	\hat{H}&=& -t \sum_{\left\langle i,j \right\rangle,\sigma} \hat{c}^\dag_{i\sigma} \hat{c}_{j\sigma}
	           -\mathrm{i}\lambda \sum_{\left\langle\left\langle i,j \right\rangle\right\rangle,\sigma} v_{ij} \hat{c}^\dag_{i\sigma} \hat{c}_{j\sigma}   \nonumber \\
	       &&   + \Delta \sum_{i,\sigma} s_i \hat{n}_{i\sigma} + U \sum_{i}\left(\hat{n}_{i\uparrow}-\frac{1}{2}\right) \left(\hat{n}_{i\downarrow}-\frac{1}{2} \right), 
	 \label {eq:Ham}
\end{eqnarray}
where $\hat{c}^\dag_{i\sigma}$ ($\hat{c}_{i\sigma}$) creates (annihilates) a fermion at site $i$ of the honeycomb lattice with spin $\sigma\in\left\{\uparrow,\downarrow\right\}$, $\hat{n}_{i\sigma}\equiv \hat{c}^\dag_{i\sigma} \hat{c}_{i\sigma}$ denotes the occupation number operator, $t$ is the hopping amplitude between nearest neighbors $\left\langle i,j\right\rangle$, and $\mathrm{i}\lambda$ is the purely imaginary hopping between next-nearest-neighbor sites $\left\langle\left\langle i,j\right\rangle\right\rangle$. $v_{ij}=-1$ ($+1$) for the hopping from site $i$ to $j$ in (anti-)clock-wise direction with respect to the center of the hexagon, illustrated in Fig.~\ref{fig:model}. The sign $s_i$ is $+1$ on one sublattice of the honeycomb lattice and $-1$ on the other.
The last term is the onsite repulsive interaction with strength $U>0$. Without loss of generality we assume $\lambda\geq 0$.

The main focus of our study is the half-filled {Haldane--Hubbard model} with $\lambda \ne 0$ and $\Delta=0$. Without interactions ($U=0$), the ground state is a topologically non-trivial Chern insulator (CI) with Chern number 1 for both spin species and a band gap $\min(\sqrt{27} \lambda, t)$. For comparison, we also consider the ionic Hubbard model on the honeycomb lattice with staggered chemical potential $\Delta\neq0$ and $\lambda=0$. In this case the non-interacting system also has a finite band gap, determined by $\Delta$, but it is topologically trivial. The full model (\ref{eq:Ham}) can be experimentally implemented with independent tunability of each term.~\cite{HaldaneRealization14, PhysRevLett.115.115303} 

\begin{figure}
\centering
\includegraphics[width=6.5cm]{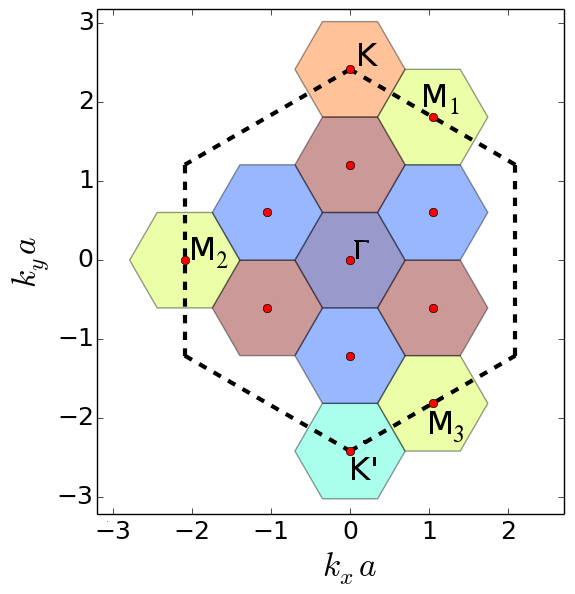}
\caption{The DCA patches in the reciprocal space for the 24-site cluster used throughout the study. The number of DCA patches, $12$, equals to the number of unit cells contained by the cluster. The Brillouin zone of the lattice is the interior of the dashed hexagon. All high symmetry points of the Brillouin zone, $\Gamma, K, K^\prime$, and the three "time-reversal symmetric" points $M_i$, are located at a patch center. The nearest neighbor distance of sites in realspace is denoted by $a$.
}
\label{fig:cluster}
\end{figure}

To map out the ground state phase diagram of Eq.~({\ref{eq:Ham}}) using the DCA method,\cite{DCA2001,QClusterTheoriesReview2005} we solve a cluster impurity problem embedded selfconsistently into a bath using continuous-time auxiliary-field QMC method with sub-matrix updates.~\cite{CTAUX2008,CTAUXsubmatrixUpdates2011} Details of the DCA method for multisite unit cells are described in Ref.~\onlinecite{StackedHoneycombAndSquare}. 
For most of this study we use the cluster shown in~Fig.~\ref{fig:model}, which respects the three-fold rotational symmetry of the honeycomb lattice. Its reciprocal representation displayed in Fig.~\ref{fig:cluster} contains all the high symmetry reciprocal lattice points of our model. The non-interacting dispersion of $\hat{H}$ is linear at $K$ and $K^\prime$ only for $\Delta=\lambda=0$. The $K$ and $K^\prime$ points remain the points of minimal non-interacting band gap for $\lambda /t \leq 1/\sqrt{27} \approx 0.192$ irrespective of $\Delta$.

For $\lambda=0$ and $\Delta=0$, the model reduces to the honeycomb lattice Hubbard model where sign-problem free QMC simulations have shown a continuous phase transition from a Dirac semi-metal to an AFI.\cite{NoSpinLiquid2012,AssaadPinningTheOrder2013} However, the model suffers from a sign problem\cite{SignProblem,SignProblemInQMC} for $\lambda\neq 0$ or $\Delta\neq 0$. Even though the sign problem is mitigated in the DCA approach compared to lattice QMC simulations, it still limits the accessible cluster size, temperature, and parameter ranges of $\lambda$ or $\Delta$. We perform simulations at a temperature $T/t=1/16$, which corresponds to the bulk non-interacting gap of the Haldane model at $\lambda/t\approx 0.012$. This temperature is below all relevant energy scales and should thus exhibit ground state behavior of the model. The sign problem limits the accessible range of $\lambda$ for the chosen cluster and temperature to $\lambda/t \leq 0.15$, which nevertheless lies in the experimentally relevant region.\cite{HaldaneRealization14}

To characterize the magnetic properties of the system we measure  the staggered magnetization in the cluster
\begin{equation}
	m = \frac{1}{N} \sum_i s_i \left\langle \hat{n}_{i\uparrow} - \hat{n}_{i\downarrow} \right\rangle,
\end{equation}
with $N$ being the number of sites of the cluster.
While the investigated two-dimensional model cannot spontaneously break the continuous symmetry at non-zero temperature,\cite{MerminWagnerTheorem1966,Hohenberg1967} the DCA solution at a low but non-zero temperature $T$ may still develop magnetic long-range order as DCA treats long-range correlations in a mean-field fashion. Such ordered solution should  be thought of as a DCA approximation of the ground state. By systematically increasing the cluster size the DCA result then becomes increasingly accurate.

To reveal the topological nature of the phase we compute the Chern number, using the topological Hamiltonian of Ref.~\onlinecite{TopoHam_ZWang_13},
\begin{equation}
	H_{\textrm{topo}}({\bf k}) \equiv -G^{-1}(i\omega=0,{\bf k})=H_0({\bf k})+\Sigma(i\omega=0,{\bf k}),
\end{equation}
where $H_0({\bf k})$ is the non-interacting part of the Hamiltonian~(\ref{eq:Ham}).
We obtain $\Sigma(i\omega=0,{\bf K})$ by a cubic spline interpolation over 40 lowest (positive and negative) Matsubara frequency self energies $\Sigma(i\omega_n,{\bf K})$. In the DCA, the self energy $\Sigma({\bf k})$ is approximated by the impurity self energy $\Sigma({\bf K})$ at the closest cluster  momentum ${\bf K}$, {\it i.e.} it is a patch-wise constant function in reciprocal space. The Chern number calculation utilizing $H_{\textrm{topo}}$ is performed by discretization of the Brillouin zone as in Ref.~\onlinecite{ChernNumberComputation_Suzuki_05}. The results are robust with respect to different Brillouin zone discretization meshes. In addition we checked robustness of the results with respect to interpolation of the self energy in reciprocal space using natural neighbor interpolation. The Chern number, being a topological invariant, may change only if the topological gap, i.e. the band gap of $H_\textrm{topo}({\bf k})$, closes. 
We find that for all examined values of $\lambda$, i.e. for $\lambda/t \leq 0.15$, the topological gap closes at the $K$ and $K^\prime$ point,\footnote{In case of the discontinuities, the topological gap closes during the run of the self-consistency, for a non-converged iteration.} 
while the single particle gap of the physical Hamiltonian (\ref{eq:Ham}) remains finite.

\section{Results}

\subsection{Phase diagram \label{sec:phasediag}}

\begin{figure}[t]
\centering
\includegraphics[width=8.5cm]{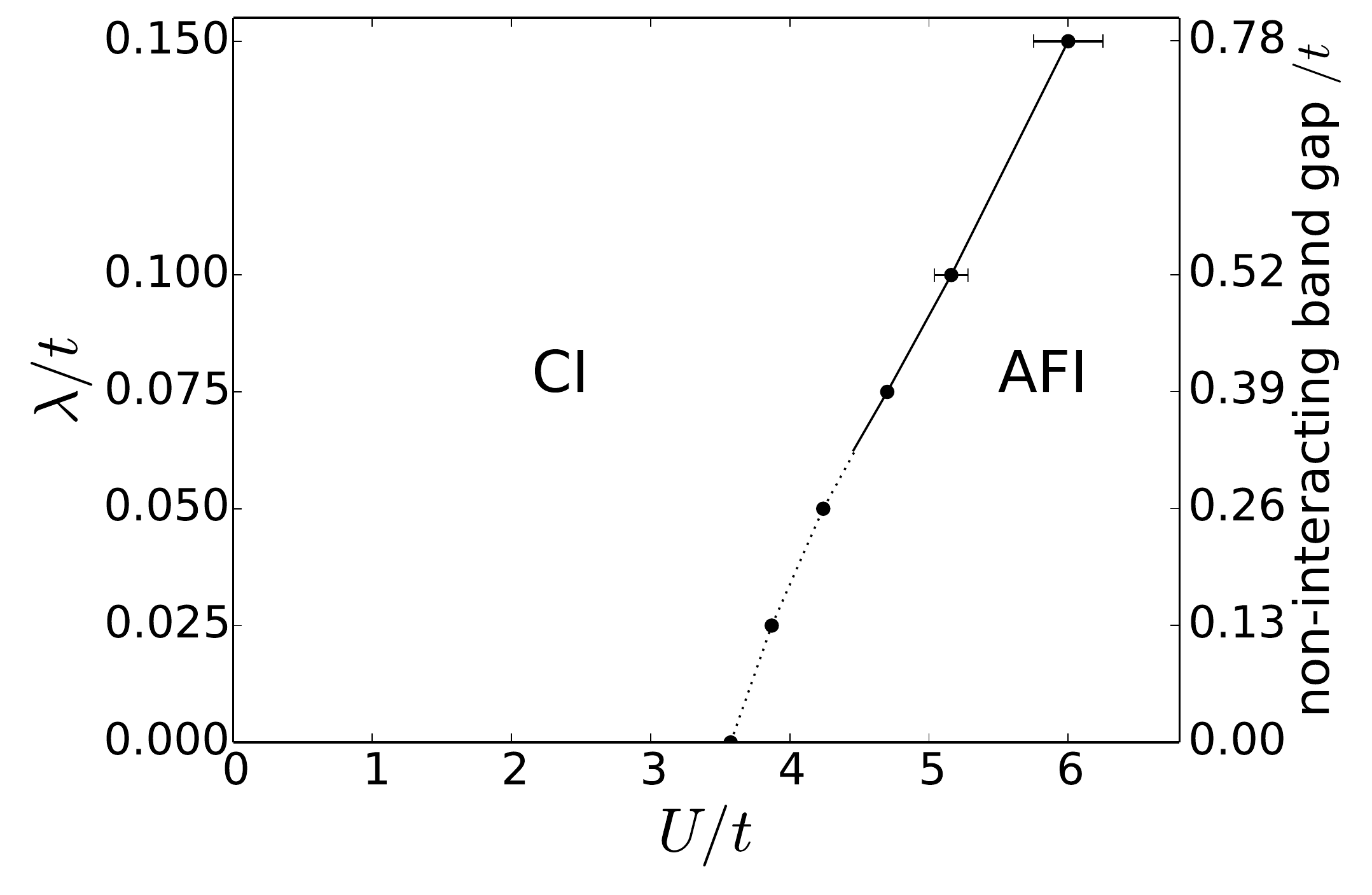}
\caption{The phase diagram of the Haldane--Hubbard model on honeycomb lattice based on simulation at $T/t=1/16$ using the 24-site cluster. The solid line is a first order phase transition in the Haldane--Hubbard model from the CI to the AFI. On the dotted line we do not have confidence about the character of the transition even though we observe continuous phase transition (see text for discussions). The error bars of the data points indicate the range of the hysteresis. 
The right vertical axis shows the size of the non-interacting band gap, $\sqrt{27}\lambda$.}
\label{fig:phase_diagram}
\end{figure}

Figure~\ref{fig:phase_diagram} shows our phase diagram of the Haldane--Hubbard model. For $\lambda/t \geq 0.075$, we find clear evidence of a first order transition from the CI to the topologically trivial AFI shown by the black solid line.
This phase boundary is not extrapolated in cluster size. To assess the systematic error, we consider the $\lambda=0$ limit where the model reduces to the honeycomb lattice Hubbard model where unbiased QMC methods predict a critical interaction to lie between $3.78t$ and $3.9t$.\cite{NoSpinLiquid2012,AssaadPinningTheOrder2013} The unextrapolated value $U_{\textrm{HH}}(\lambda=0)/t=3.575 \pm 0.075$ based on our 24-site cluster underestimates this value by about $0.3t$, as the DCA transition occurs when the correlation length reaches the order of the cluster size. This difference provides an estimate of the systematic error. At the first order transition the systematic error is expected to be smaller. 
For $\lambda / t \leq 0.05$ DCA with 24-site cluster is consistent with a continuous phase transition with intermediate topological AFI. However we believe this to be due to insufficiently large cluster and that using larger clusters will again lead to a first order transition.

\subsubsection{First order transition for $\lambda/t \geq 0.075$}
\label{subsec:large_lamdba}

\begin{figure}[b]
\centering
\includegraphics[width=8.5cm]{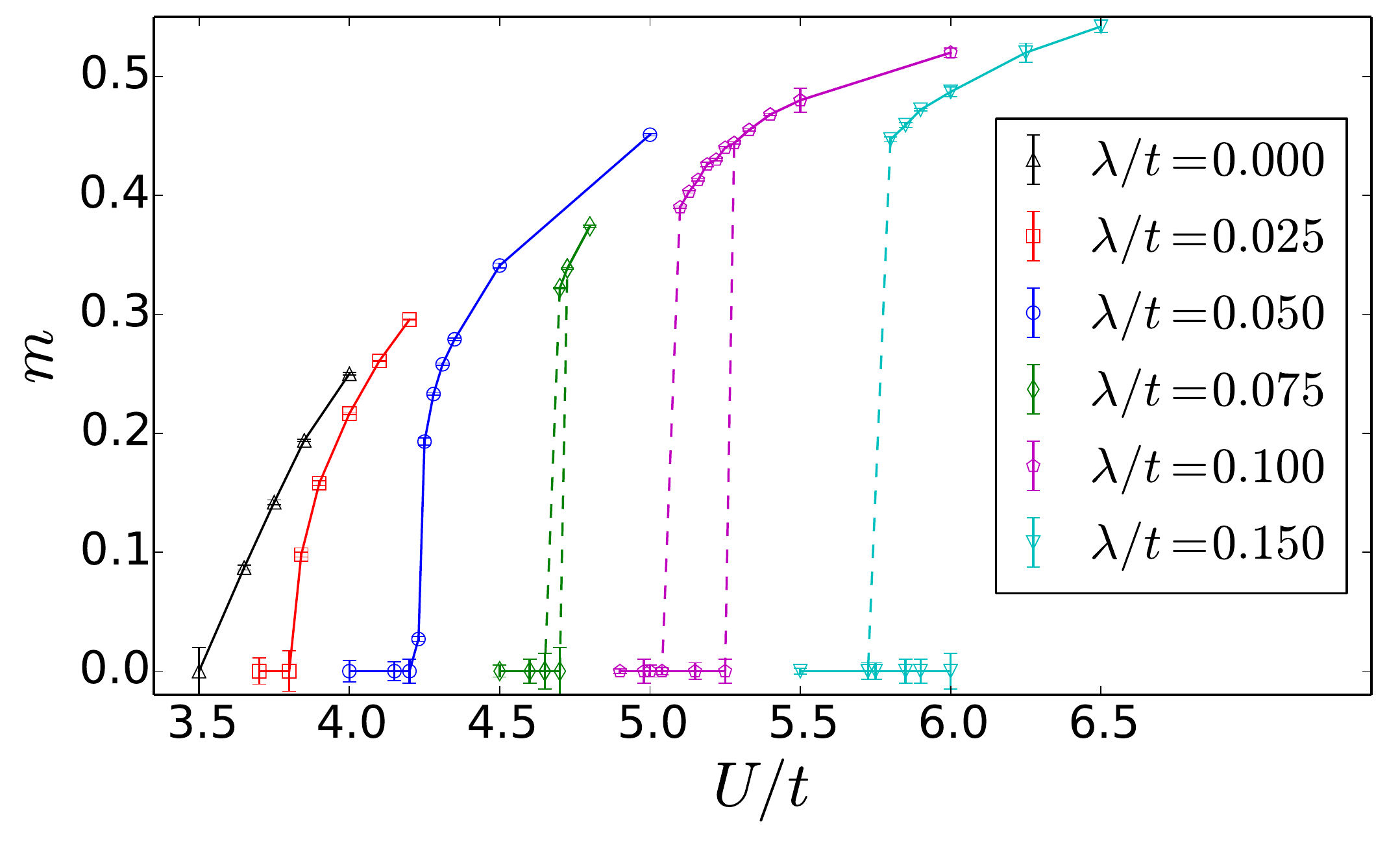}
\caption{The staggered magnetization as a function of $U/t$ for different $\lambda/t$ obtained for a 24-site cluster at $T/t=1/16$. The dashed lines indicates a  discontinuity of the staggered magnetization, and the region between the dashed lines indicates the hysteretic region where it is possible to converge to either a paramagnetic or an antiferromagnetically ordered solution. This hysteresis is visible for $\lambda/t=0.075$, $0.1$, $0.15$. For $\lambda=0.15t$ we do not provide the upper bound for stability of the paramagnetic phase due to a too large  sign problem.}
\label{fig:magnetization}
\end{figure}

Figure~\ref{fig:magnetization} shows the  staggered magnetization as a function of  $U/t$ for various values of $\lambda$.
Noticeably, $m$ shows a discontinuity for $\lambda/t \geq 0.075$, accompanied by hysteretic behavior. The simulation can converge to two different solutions depending on the initial bare cluster Green's function provided to the self-consistency loop. This provides a clear signature of a first order phase transition at $\lambda/t \geq 0.075$. 
In order to distinguish between slow convergence of the self-consistency procedure and (meta)stable solutions we perform about one hundred iterations.

\begin{figure}
\centering
\includegraphics[width=8.5cm]{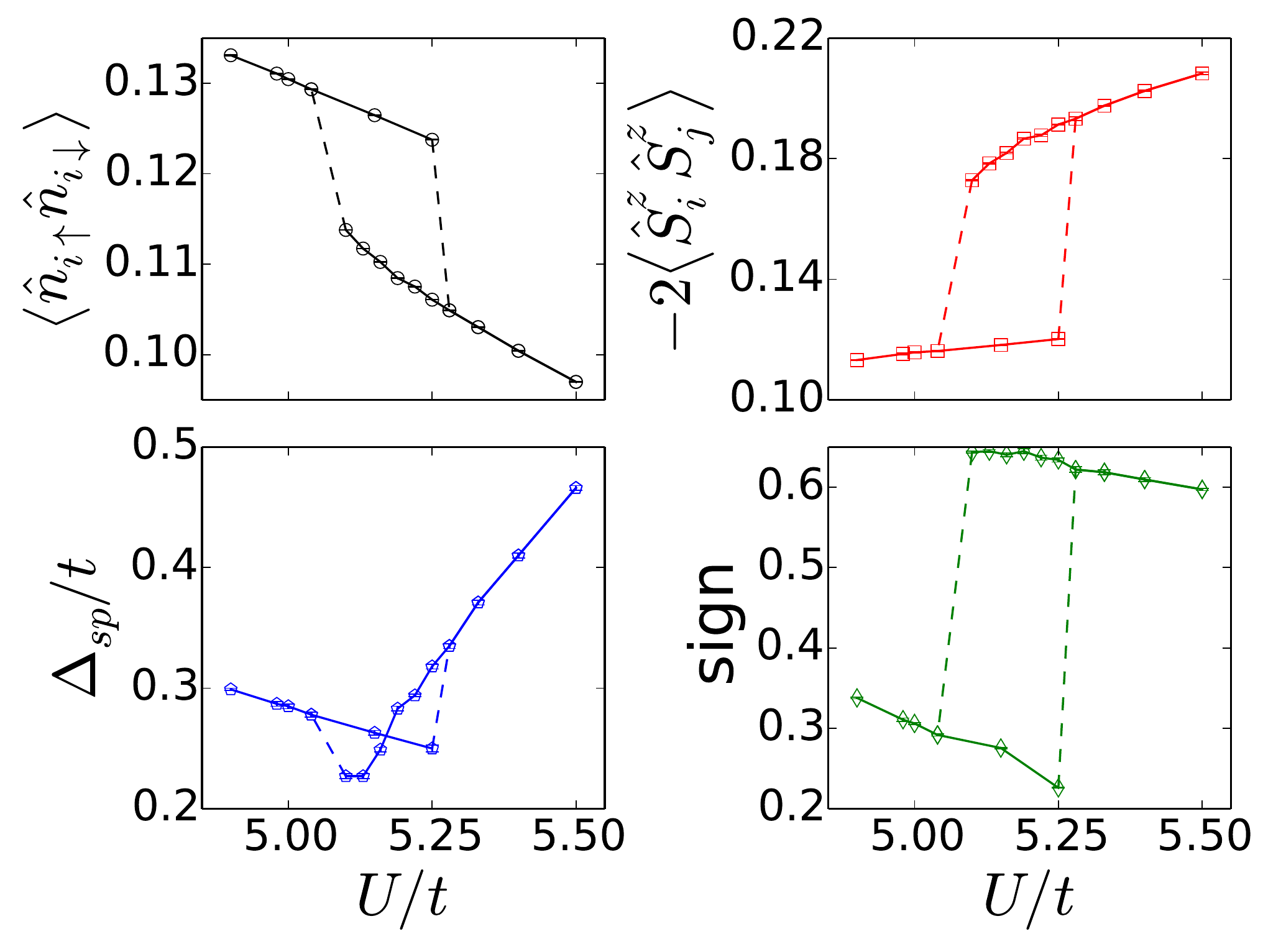}
\caption{Double occupancy $\left\langle \hat{n}_{i\uparrow} \hat{n}_{i\downarrow} \right\rangle$,  nearest neighbor spin-spin correlation function $-2 \left\langle \hat{S}^z_i \hat{S}^z_j \right\rangle$, single particle gap $\Delta_{\textrm{sp}}$, and the average sign in the simulation are shown as a function of $U$ at $\lambda/t=0.1$. The dashed line segments show the discontinuity at the first order phase transition.}
\label{fig:lambda0.1}
\end{figure}

Around the first order transition other observables also exhibit hysteretic behavior, as shown in Fig.~\ref{fig:lambda0.1} for $\lambda/t=0.1$. The two curves in each panel are obtained with self-consistent iterations  started either from the CI or from the AFI state.
Hysteretic behavior in these observables can also be  measured experimentally as a signature of a first order phase transition. The single particle gap $\Delta_{\textrm{sp}}$ is obtained from the imaginary time lattice Green's function at the $K$ point by fitting to $A\cosh\left[\Delta_{\textrm{sp}}(\tau-B)\right]$ near $\tau=1/(2T)$. Note that in this case the temperature $T/t=1/16$ is for all values of $U$ at least four times smaller than the single particle gap, and thus low enough to capture ground state behavior.
The average sign of the impurity solver is also notably different in the two phases and exhibits a jump at the transition point. The Chern number (not plotted) equals to $1$ for the non-magnetic solutions and to $0$ for the magnetic solutions.

This clear evidence of a first order phase transition is different than the continuous phase transition transition found in the static mean-field~\cite{SlaveBoson, PhysRevB.84.035127, HaldaneHubbardMeanField_2015, PhysRevB.83.205116, HaldaneHubbardMeanField_2015b} and two-site cellular DMFT~\cite{HaldaneHubbard_QClusterSenechal_2015} (CDMFT) studies. Since our DCA calculation on a 24-site cluster incorporates  short-range correlation effects and we can reproduce some of the continuous transition character by using small clusters (see Sec.~\ref{subsec:comparison}), we believe the first order transition found in the Haldane--Hubbard model is real.


\subsubsection{Phase transition for $\lambda/t < 0.075$}
\label{subsec:small_lambda}

\begin{figure}
\centering
\includegraphics[width=8.5cm]{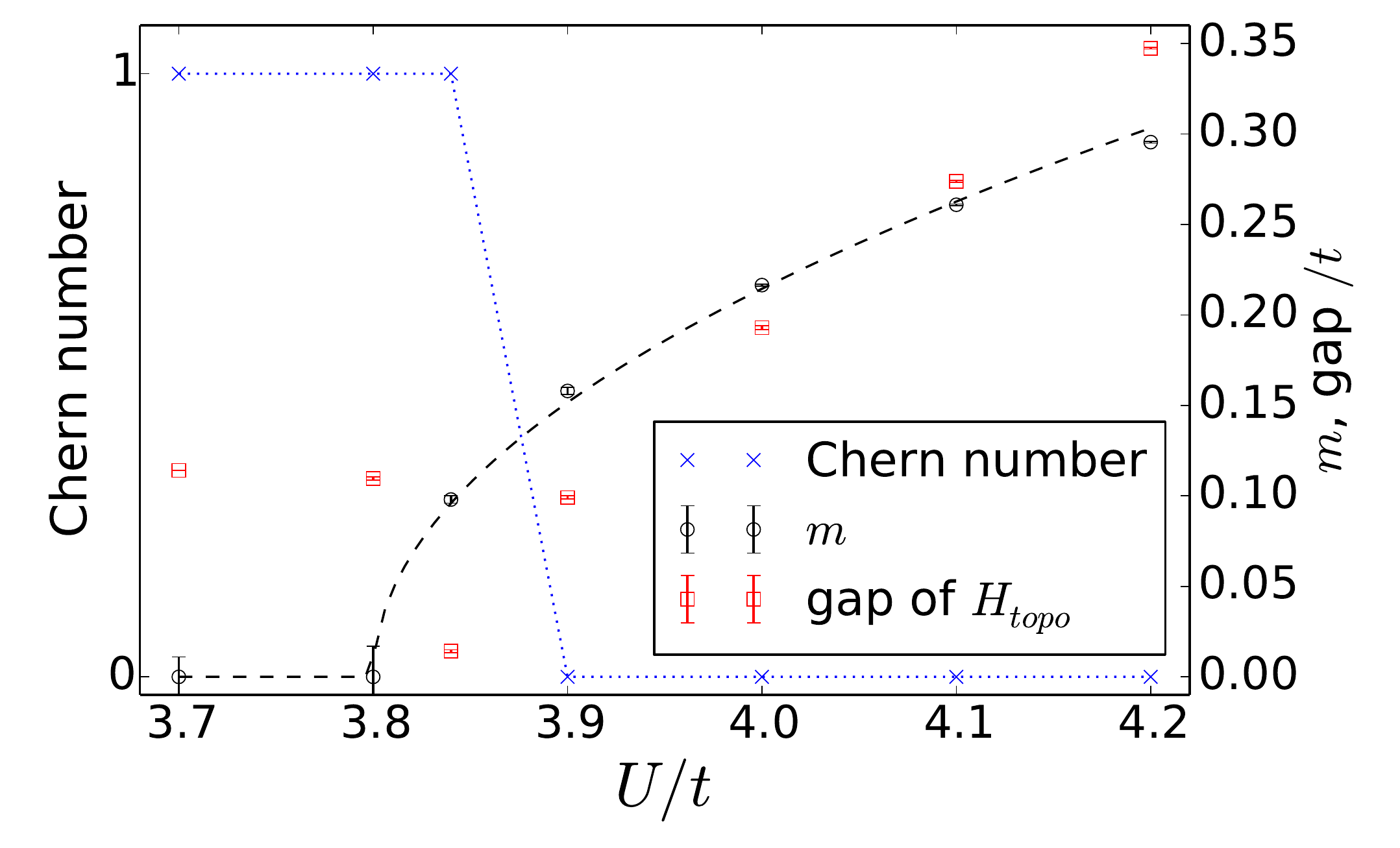}
\caption{The staggered magnetization and the topological gap at the $K$ point as a function of $U/t$ for $\lambda/t=0.025$. The fit of the staggered magnetization in range $U/t \in [3.7, 4.1]$ uses the mean-field critical exponent $\beta_{\textrm{mf}}=0.5$. The Chern numbers of the occupied bands drops between $3.84<U/t<3.9$, consistent with closing of the gap of $H_\textrm{topo}$ in the same range. The dotted line is a guide to the eye.}
\label{fig:lambda0.025}
\end{figure}

For $\lambda/t \leq 0.05$, we find a continuous increase of the staggered magnetization, as shown in Fig.~\ref{fig:magnetization}. As a consequence of the smooth increase of the magnetic order parameter, an intermediate topologically non-trivial AFI appears in between the CI for low $U$ and the AFI for large $U$. The simulation results at $\lambda/t=0.025$ are depicted in terms of the topological gap at the $K$ point, the staggered magnetization $m$, and the Chern number as a function of $U$ in Fig.~\ref{fig:lambda0.025}. The Chern number drops from $1$ to $0$ inside the magnetic ordered phase. The same scenario is found at $\lambda/t = 0.05$.

Even though our DCA results are consistent with an intermediate topological AFI state in a small region of parameter space, the  data is also consistent with the scenario of a first order phase transition for any non-zero $\lambda$, and a diverging correlation length as $\lambda \rightarrow 0$. In this scenario, the correlation length at the first order transition remains finite at any non-zero $\lambda$, but is larger than the 24-site cluster employed here, thus resulting in an apparent continuous phase transition for $\lambda/t \leq 0.05$ region.
Larger clusters would thus be required to resolve the phase transition character in the small $\lambda$ region, but are intractable because of the sign problem.

Finally, for $\lambda = 0$, the model reduces to the Hubbard model on the honeycomb lattice, where there is a firm evidence that the  model undergoes a direct second order phase transition from the paramagnetic semimetal to the AFI.\cite{NoSpinLiquid2012,AssaadPinningTheOrder2013}
A DCA study of that model predicts in agreement with the latter studies the direct second order phase transition.\cite{DCAHoneycomb}
In the vicinity of a second order phase transition the correlation length exceeds the cluster size and then mean-field behavior appears in the DCA solution.

\subsection{Comparison to the ionic Hubbard model on the honeycomb lattice}
\label{subsec:ionic_honeycomb}

\begin{figure}
\centering
\includegraphics[width=8.5cm]{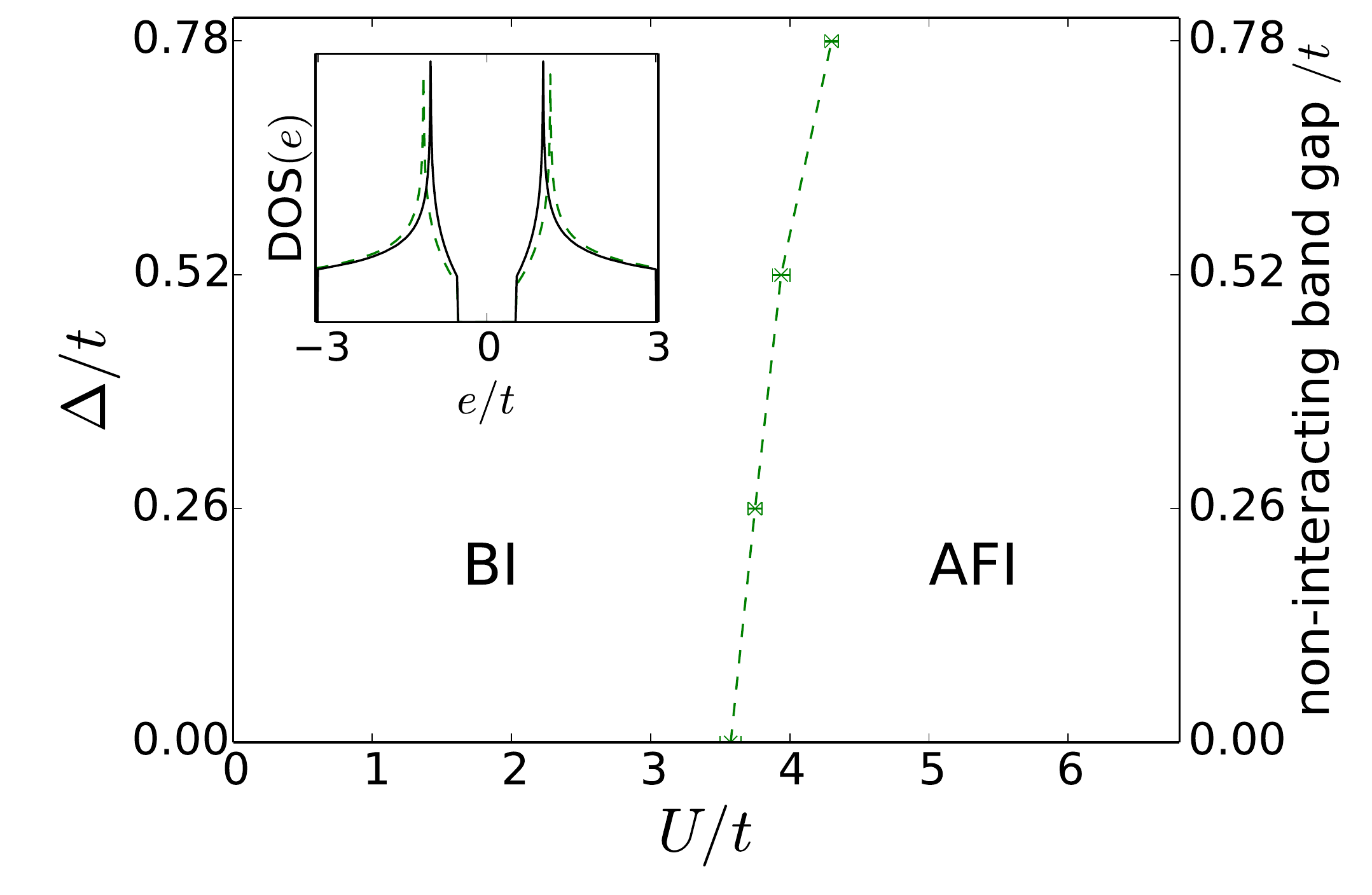}
\caption{The phase diagram of the ionic Hubbard model on the honeycomb lattice based on simulations at $T/t=1/16$. The dashed line denotes the  critical interaction strength $U_{\textrm{IH}}$ of a  second order phase transition from a band insulator to an AFI. 
The error bars show the bounds for the  onset of ordering for a 24-site cluster.
The inset shows the density of states of the non-interacting Haldane model at $\lambda/t=0.1$ as a solid black line, and that of the non-interacting ionic model for $\Delta=0.52 t$ as a dashed green line. Both models have a band gap of $0.52 t$.}
\label{fig:phase_diagram_ionic}
\end{figure}

To further reveal the role of the topological band gap, we compare the phase diagram of the Haldane--Hubbard model with that of the ionic Hubbard model on honeycomb lattice. The latter model is defined by $\lambda=0$ and a staggered sublattice potential $\pm\Delta$, which opens a topologically trivial band gap. The non-interacting dispersion of the Haldane and the ionic models are similar if the non-interacting band gaps are adjusted to match each other. The density of states for both models with non-interacting band gap $0.52 t$ is shown in the inset of Fig.~\ref{fig:phase_diagram_ionic}.\footnote{The largest difference between the non-interacting dispersions of the two compared models is at the $M$ points, $\sqrt{t^2-27\lambda^2}-t$, where this is the shift of the van Hove peaks in the non-interacting density of states.} Thus, in a crude theoretical treatment which only cares about the band gap or density of states, these two models should have similar phase diagrams. However, the phase diagram of the ionic honeycomb model shown in Fig.~\ref{fig:phase_diagram_ionic} differs substantially from that of the Haldane--Hubbard model in Fig.~\ref{fig:phase_diagram}. The dependence of the critical interaction strength on the non-interacting band gap in the ionic model is weaker than in the Haldane--Hubbard model. More importantly, the character of the transition is second order in the ionic Hubbard model for all simulated parameters. 

While the ionic Hubbard model on the square lattice exhibits an intermediate metallic phase between the ionic band insulator (BI) and AFI,
\cite{IonicHubbard_Scalettar_07,IonicHubbard_DMFT_Dagotto_07,IonicHubbard_Scalettar_07b}
our simulations find no indication of such phase on the honeycomb lattice. A reason for this difference may be different position of the van Hove singularities, which are at the band edges for the square lattice, but not for the honeycomb lattice. A similar observation was made in Ref.~\onlinecite{IonicHoneycomb_CMDFT_15}. 

\subsection{Comparison to small cluster calculations}
\label{subsec:comparison}

Both static mean-field calculations\cite{SlaveBoson, PhysRevB.84.035127, HaldaneHubbardMeanField_2015, PhysRevB.83.205116, HaldaneHubbardMeanField_2015b} and CDMFT on 2-site clusters~\cite{HaldaneHubbard_QClusterSenechal_2015} predict a continuous phase transition from CI to the AFI, with an intermediate topologically non-trivial AFI phase for a wide range of $\lambda$.
Also results of a variational cluster approximation calculation on 6-site clusters \cite{HaldaneHubbard_QClusterSenechal_2015,HaldaneHubbard_VCA_2015} indicate an indirect transition from CI to AFI, but via a topologically non-trivial non-magnetic insulating phase with opposite Chern number as the CI.
Another recent study using 6-site DCA and CDMFT calculations\cite{HaldaneHubbard_DMFT_ED_2015} reports, similar to our findings, a signature of a first order transition for $\lambda/t=0.2$. Since all studies mentioned above employ quantum cluster approaches of a similar nature, these discrepancies may either be due to insufficiently large clusters or due to subtleties in the cluster embeddings which break the spatial symmetries.\cite{PhysRevLett.111.029701, PhysRevLett.111.029702}  

\begin{figure}
\centering
\vspace{4mm}
\includegraphics[width=2.39cm]{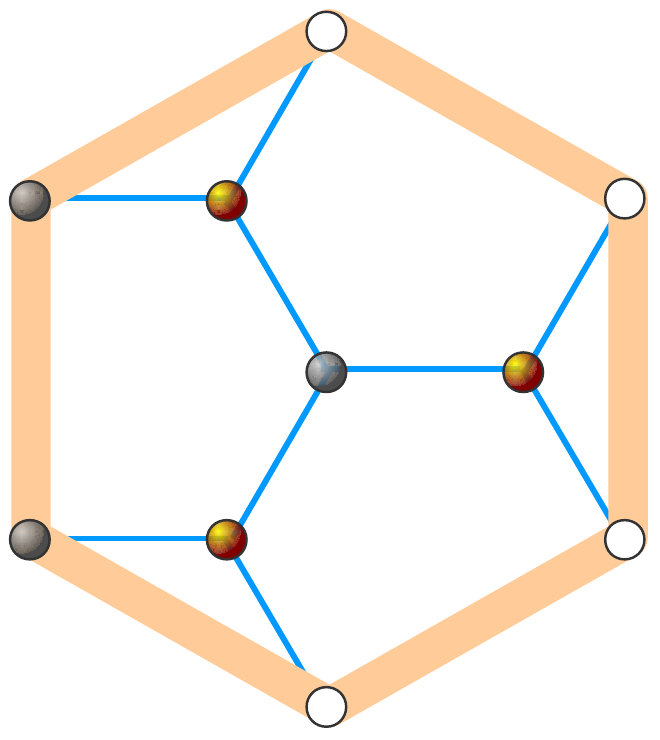}
\hspace{10mm}
\includegraphics[width=3.11cm]{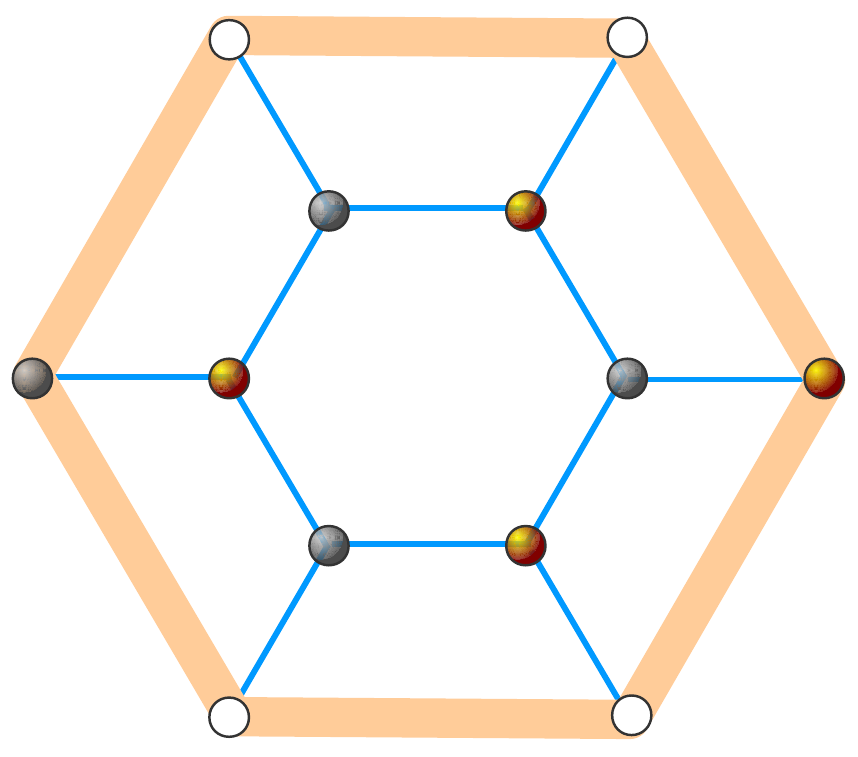}
\includegraphics[width=7.0cm]{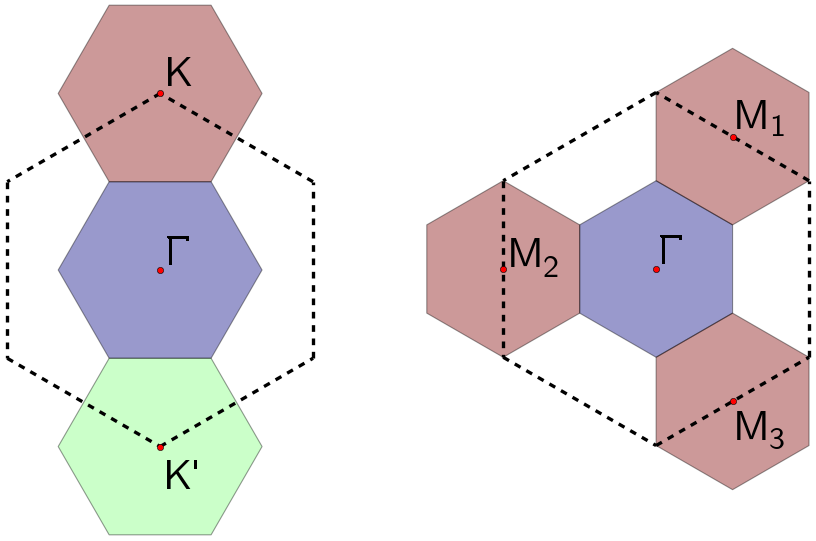}
\caption{Top: The 6-site  and the 8-site cluster shown in realspace. As in the Fig.~\ref{fig:model}, the white sites on the border correspond via periodic boundary conditions to the border sites. Bottom: The DCA patches for the 6-site  and the 8-site clusters  in reciprocal space. The Brillouin zone of the lattice is the interior of the dashed hexagon. The 6-site cluster contains the $K$, $K^\prime$ points as patch centers, while the 8-site cluster does not.
}
\label{fig:small_clusters}
\end{figure}

\begin{figure}
\centering
\includegraphics[width=8.5cm]{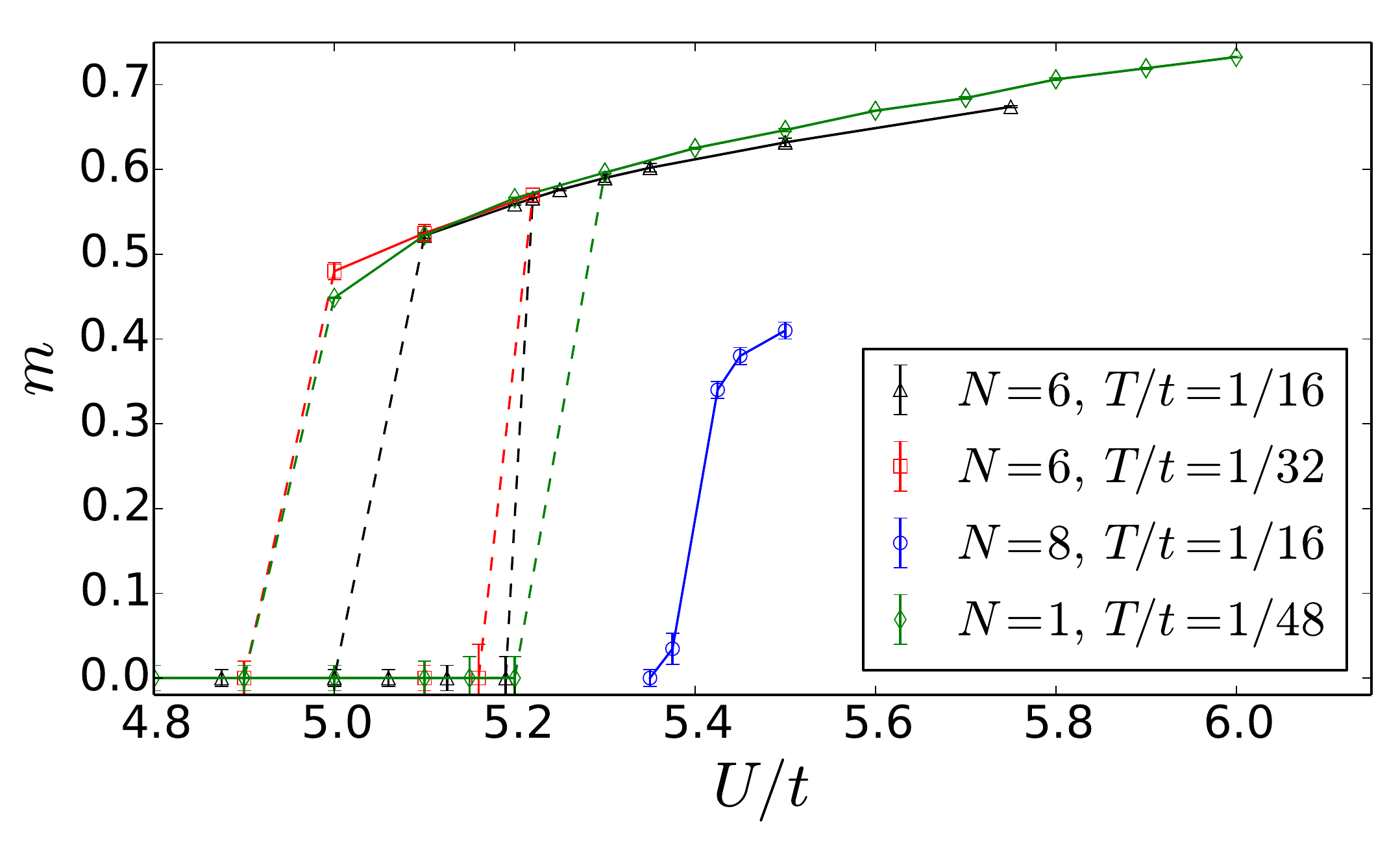}
\caption{The staggered magnetization obtained for the Haldane--Hubbard model at $\lambda/t=0.1$ using additional small clusters shown in Fig.~\ref{fig:small_clusters} simulated at temperature $T$. The single-site DMFT data ($N=1$) is presented as well.}
\label{fig:magnetization_different_clusters}
\end{figure}

To shed light on this issue we examined the Haldane--Hubbard model using two additional clusters of different size, shown in Fig.~\ref{fig:small_clusters}. The 6-site cluster contains the $K$ and $K^\prime$ point in its reciprocal representation, while the 8-site cluster does not. Both of them respect the three-fold rotational symmetry. The staggered magnetization obtained using these clusters at $\lambda/t=0.1$ are shown in Fig.~\ref{fig:magnetization_different_clusters}. The 6-site cluster displays similar hysteresis as  observed above for the 24-site cluster, with a difference in the transition point $U/t$ of at most $0.1$. The value of $m$ in the ordered phase is larger than for the 24-site cluster, which is expected, as DCA becomes exact for $N \rightarrow \infty$ and $m$ has to vanish in the thermodynamic limit at $T\ne0$. In contrast, using the 8-site cluster we observe a sharp but continuous increase of $m$ at a strongly shifted transition point $U_{\textrm{HH}}/t=5.40 \pm 0.03$ for $\lambda/t=0.1$, without any trace of hysteresis. These findings are similar to those obtained for Haldane model of spinless fermions,\cite{InteractingSpinlessHaldane10,InteractingSpinlessHaldane10b,InteractingSpinlessHaldane11} where Varney \emph{et al.}, using exact diagonalization, observed first order or continuous transition character depending on the presence of the $K$ and $K^\prime$ points in the cluster reciprocal representation. Our findings in DCA support their conclusion that the choice of the cluster is significant and that reliable clusters need to contain the $K$ and $K^\prime$ points.

Examining the model within the DMFT approximation yields further insight. For this we simulate a single site in sublattice $A$. The form of the ${\bf k}$-independent DMFT self energy for a single unit cell (containing two sites) is obtained only from the self energy of the simulated site. Its form has to respect the symmetry of the studied Hamiltonian~\ref{eq:Ham}, which is invariant (up to irrelevant constants) under particle-hole transformation combined with spatial inversion,
ensuring 
\begin{equation}
	G_{AA\sigma}(i\omega_n, {\bf k}) = -G_{BB\sigma}^*(i\omega_n, {\bf k}) \,,
\end{equation}
 both in the  paramagnetic and in the  antiferromagnetically ordered phase. The DMFT self energy is then approximated by
\begin{eqnarray}
	\Sigma_{BB\sigma}(i\omega_n) = -\Sigma_{AA\sigma}^*(i\omega_n) \,, \label{eq:DMFT_self energy_Ansatz1} \\
	\Sigma_{AB\sigma}(i\omega_n)=0=\Sigma_{BA\sigma}(i\omega_n) \,, \label{eq:DMFT_self energy_Ansatz2}
\end{eqnarray}
neglecting the $AB$ components, motivated by the dominantly local character of the self energy for a Hubbard interaction. 
The DMFT mapping is then formulated conveniently with $2\times 2$ matrices, comprising the sublattice indices,
\begin{equation}
	G_{AA\sigma}(i\omega_n) = \frac{1}{\Omega} \int_{\textrm{BZ}} \mathrm{d}{\bf k}\: \left[ \left(G_{\sigma}^{0}(i\omega_n,{\bf k})\right)^{-1} -\Sigma_\sigma(i\omega_n) \right]_{AA}^{-1} \,, \label{eq:DMFT_mapping}
\end{equation}
 integrating over the Brillouin zone (with volume $\Omega$). Here $G_\sigma^0(i\omega_n, {\bf k})=\left[i\omega_n \mathbb{1}-H_0({\bf k})\right]^{-1}$ is the non-interacting lattice Green's function.
Surprisingly, the magnetization curve for the Haldane--Hubbard model at $\lambda/t=0.1$ shows discontinuities and hysteresis (Fig.~\ref{fig:magnetization_different_clusters}) even in the DMFT simulation. This apparent contradiction to our conclusion about the necessity of the $K$ and $K'$ point in the cluster reciprocal representation can be explained by the prescribed form of the DMFT self energy in Eqs.~(\ref{eq:DMFT_self energy_Ansatz1}-\ref{eq:DMFT_self energy_Ansatz2}), which \emph{coincidentally} obeys the same constraints, of vanishing $AB$ components, as those due to the symmetry of the self energy at the $K$ and $K'$ point, arising from the three-fold rotational symmetry of the model.
For the ionic honeycomb model simulated by DMFT at $\Delta/t=0.52$, $m$ is continuous. Note that the next-nearest neighbor hoppings on the same sublattice ($\lambda\neq 0$) do not allow one to rewrite the mapping~Eq.~(\ref{eq:DMFT_mapping}) as an integral over the density of states, which explains the possibility of finding qualitatively different behaviors of the magnetization in the Haldane--Hubbard and the ionic honeycomb model despite their very similar non-interacting density of states.

Finally, the 6-site cluster enable simulations at lower temperature since the sign problem is less severe than for the 24-site cluster. Results obtained at twice lower temperature differ only by an enlarged ordered phase (see Fig.~\ref{fig:small_clusters}), while the first order characteristics remain unchanged.

\section{Outlook}
\label{sec:discussion}

Our predictions can be checked by the experiments on the Haldane--Hubbard model in optical lattice simulators.~\cite{PhysRevLett.115.115303,HysteresisInOpticalLattice_Esslinger_15} The first order phase transition can be detected as a hysteresis of spatially averaged local observables. By tuning the interaction strength to the coexisting region one may also find coexisting domains of CI and  AFI phases. Each AFI domain is of the size of the magnetic correlation length at the first order transition point. Interestingly, the topological nature of the CI would imply presence of chiral edge states around the domain walls which may be revealed by an in-situ measurement of the domains in the ultracold atomic gas. 

\begin{acknowledgments}
We thank Emanuel Gull for providing us with the continuous-time auxiliary-field impurity solver for real Hamiltonians,~\cite{CTAUX2008} which we adapted to handle complex Hamiltonians. The code is based on the ALPS libraries\cite{ALPS1.3:2007,ALPS2.0:2011} and calculations were performed on the M\"onch cluster. This work was supported by the European Research Council through ERC Advanced Grant SIMCOFE and by the Swiss National Science Foundation through NCCR QSIT.
\end{acknowledgments}

\bibliographystyle{apsrev4-1}
\bibliography{haldane_hubbard,haldane_hubbard_lw}

\end{document}